\documentclass[
rsi,
 amsmath,amssymb,
preprint,
]{revtex4-1}

\usepackage{graphicx}
\usepackage{dcolumn}
\usepackage{bm}

\usepackage[utf8]{inputenc}
\usepackage[T1]{fontenc}
\usepackage{mathptmx}
\usepackage{etoolbox}
\graphicspath{{Figures/}}
\usepackage{svg}
\usepackage[colorlinks=true, linkcolor=blue, citecolor=blue, urlcolor=blue]{hyperref}

\makeatletter
\def\@email#1#2{%
 \endgroup
 \patchcmd{\titleblock@produce}
  {\frontmatter@RRAPformat}
  {\frontmatter@RRAPformat{\produce@RRAP{*#1\href{mailto:#2}{#2}}}\frontmatter@RRAPformat}
  {}{}
}%
\makeatother
\begin{document}

\preprint{AIP/RSI}

\title[Development of Thomson parabola spectrometer for diagnostics of
ions driven by ultrahigh intensity laser: Simulation and numerical analysis]{Development of Thomson parabola spectrometer for diagnostics of
ions driven by ultrahigh intensity laser: Simulations and numerical analysis}
\author{Kavil Mehta$^{1*}$}
\author{Jatin Parashar$^{1}$}%
\author{Shivangi Bidoliya$^{2}$}
\author{Prashant Kumar$^{1}$}
\author{Hitesh Adalja$^{1}$}
\author{Mohammad Tayyab$^{2}$}
\author{Anand Moorti$^{2}$}
\author{Juzer Ali Chakera$^{2}$}

\affiliation{ 
$^{1}$Physical Research Laboratory, Ahmedabad, India 
}%
\affiliation{%
$^{2}$Raja Ramanna Centre for Advanced Technology, Indore, India
}%
 \email{kavil@prl.res.in, kavilm1996@gmail.com}
\date{\today}

\begin{abstract}
A Thomson parabola ion spectrometer (TPIS) has been designed and developed for diagnostics of laser-accelerated ion beams in the MeV energy range. The TPIS has been validated by ion acceleration experiment at a petawatt laser facility. Necessary simulations to evaluate the electric and magnetic field distributions have been performed with the help of a numerical simulation software to aid the selection of the spectrometer geometry while minimizing fringe-field effects. Analytical dispersion expressions have been formulated from the simulations that take into account the spatial variation in the electromagnetic field profiles. The ion deflections obtained from these expressions demonstrate an improved agreement with experimentally measured proton trajectories. The TPIS has been developed with provision to change the field-free drift region, showcasing flexibility to be employed over a broad energy range and with different experimental setups. The TPIS hence fabricated in-house has been subsequently used for detecting laser-accelerated ion beams from  aluminum (Al) foil targets of different thickness. These experiments have demonstrated the capability of the spectrometer to resolve multiple ion species with sufficient separation between them. The developed TPIS provides a compact, flexible and accurate diagnostic for high-energy laser-plasma experiments.
\end{abstract}

\maketitle

\section{Introduction}

Laser-based ion acceleration has been extensively studied in the past two decades as an alternative to the conventional radio frequency (RF) accelerators \cite{Daido_2012, BORGHESI_2014, Badziak_2018}. Compact dimensions coupled with extremely high field gradient gives laser-based accelerators an edge over RF accelerators \cite{malka_2008, hooker_2013}. The mechanism of ion acceleration depending on the intensity of laser are target normal sheath acceleration (TNSA) \cite{Macchi_2013, Passoni_2010}, Radiation pressure acceleration (RPA) \cite{Wang_2024, kim_2016}, break-out afterburner(BOA) \cite{kim_2016}, shock acceleration (SA) \cite{Pak_2018, Marcowith_2016}, etc. TNSA is the mechanism which has been pursued as it occurs at relatively low intensities of laser (10$^{17}$ – 10$^{20}$ W/cm$^2$ ). Moreover, primary usage of few micron thick targets makes handling and target preparation easier compared to RPA or BOA which require ultrathin targets. Laser contrast requirement also increases with such targets, which makes TNSA realization easier. In the TNSA mechanism, laser-generated relativistic hot electrons establish a transient sheath field at the rear surface of the target, producing electrostatic field of the order of TV/m, which accelerate protons and heavier ions from the contamination layer of the target rear surface \cite{BORGHESI_2014, roth_2016}.

Characterization of laser-accelerated ions requires determination of ion energy, species, and charge state. Diagnostic techniques such as radiochromic film (RCF) stack, Columbia resin number 39 (CR-39) nuclear track detector, time-of-flight, etc. have been employed for ion diagnostics \cite{Bolton_2014, Scuderi_2018}. Among these techniques, the Thomson parabola ion spectrometer (TPIS) has become the standard diagnostic because it provides single-shot discrimination of ion species over a broad energy range \cite{Alejo_2016, Tata_2017}. A typical TPIS consists of a pinhole, parallel electric and magnetic fields through which the ions traverse, and a detector. The pinhole allows only a small, collimated portion of the diverging ion beam to transmit while suppressing the X-rays and the ions propagating outside the acceptance angle of the spectrometer. The accelerated ion beam is deflected by TPIS based on its energy (velocity) and charge-to-mass ratio (q/m) by the fields \cite{Giorgio_2020}. These deflected ions are then recorded on a detector such as multi-channel plate (MCP), CR-39 detector, scintillators, etc. The same ion species of different energy will trace a parabola on the detection plane due to parallel electric and magnetic field directions.

Several TPIS configurations have been reported for laser-plasma experiments, with designs optimized for specific experimental conditions and energy ranges. A. Kurmanova et al. developed TPIS to study protons and alpha particles and distinguish them \cite{Kurmanova_2023}. Gwynne et al. demonstrated that optimization of electric field intensity and length can help in retaining high and low-energy ions and also contribute to the separation of ion species at the high-energy end \cite{gwynne_2014}. Jung et al. developed TPIS for dispersion of high-energy C$^{6+}$ ions and protons at high resolution (< 5\%) \cite{jung_2011}. A. Huber et al. developed a modular TPIS with scintillators as detectors suitable for high repetition rate laser facilities. They simulated the spectrometer in GEANT4 to establish a dispersion relation for each ion species \cite{Huber_2025}. These studies demonstrate that TPIS design is strongly dependent on the intended experiment of deployment. Consequently, spectrometers developed for one facility cannot be directly adopted for another without redesigning the field configuration and mechanical geometry. 

In the present study, we report the design, simulation, fabrication, and experimental validation of a TPIS in which the field-free region can be tuned as per the intended ion energy range. The design of TPIS for such a case is relatively simple because localized ionization results in a quasi-directional source which can be efficiently sampled by the TPIS. However, it is important to ensure that the traversing beam experiences a uniform field to avoid distortion in the detector plane. The preliminary design parameters were established using analytical calculations and subsequently optimized through finite-element simulations using COMSOL Multiphysics \textregistered{} \cite{comsol}  software. We have verified the electric and magnetic field distributions while minimizing fringe-field effects. Instead of assuming a constant field inside the TPIS, we have taken into account the gradual rise and fall of electric and magnetic fields and have formulated analytical expressions that are in agreement with our experimental observations. The spectrometer has been fabricated in-house using the inputs from simulations. The magnets and electrodes are housed in a mild steel (MS) enclosure, resulting in improvement of both the uniformity and intensity of magnetic field. Another salient feature of the fabricated TPIS is that it is designed to tune the field-free region, thereby making it possible to measure low or high-energy ions, as per the need of experiment. The fabricated spectrometer has been employed as diagnostics for laser-driven ion acceleration experiment at petawatt laser facility, Laser plasma division, Raja Ramanna Centre for Advanced Technology (RRCAT), Indore, for experimental validation. The design methodology, numerical optimization, fabrication, and experimental performance of the developed TPIS are presented and discussed in detail.

\section{Numerical design of Thomson parabola ion spectrometer}

\begin{figure}
\includegraphics[scale=0.4]{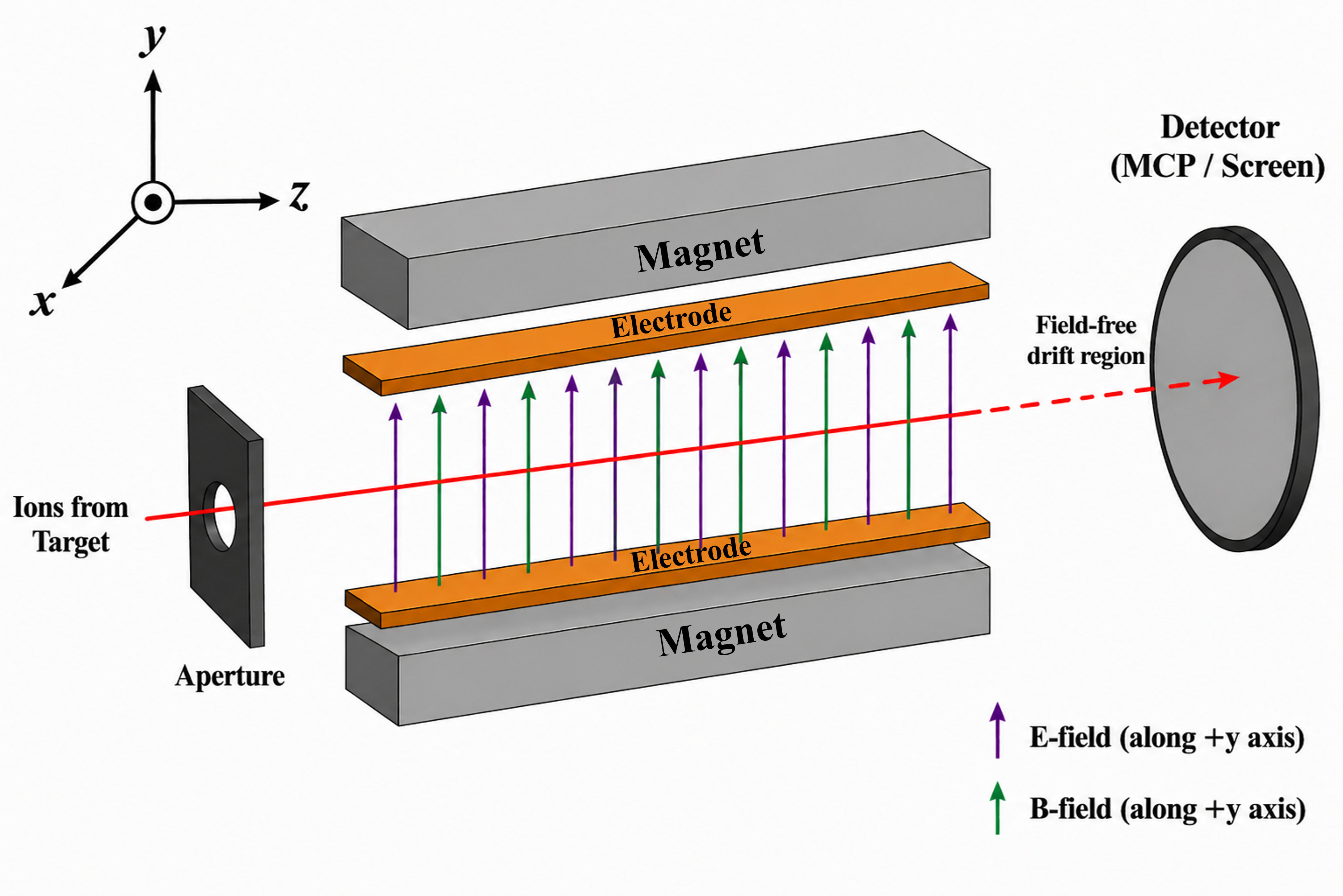}
\caption{\label{fig:fig1} Basic schematic diagram of Thomson parabola spectrometer. The magnets and electrodes have been coaxially mounted and the electric and magnetic fields are in the direction of $y$ axis.}
\end{figure}

A typical schematic of TPIS is shown in Fig. \ref{fig:fig1}. The ions traverse through pinhole into the field region. The electric and magnetic fields influence the path of ions and deflect them from their original trajectory. After the field length, the ions pass through field-free drift region, which makes the deflection more prominent before they encounter the detector. The deflection due to magnetic field is along $x$ axis (coordinate system given in Fig. \ref{fig:fig1}) and that due to electric field is along $y$ axis. The $x$ and $y$ deflections, assuming constant fields, are given as \cite{dalui_2014},

\begin{eqnarray}
x = \frac{qBL_f}{mv_z}(L_d + \frac{L_f}{2})
\label{eqn:xdef0}
\end{eqnarray}

\begin{eqnarray}
y = \frac{qEL_f}{mv_z^2}(L_d + \frac{L_f}{2})
\label{eqn:ydef0}
\end{eqnarray}
where $q$ is the charge state, $B$ and $E$ are magnetic and electric field, respectively, $m$ is the mass, $v_z$ is the velocity of ion in z-direction, $L_f$ is the field region length and $L_d$ is the drift region length.

Equations \ref{eqn:xdef0} and \ref{eqn:ydef0} consider that the fields are constant in space and time. In actual scenario, there are fringe effects along the field length, which result in a gradual rise and fall of the fields at the entrance and exit of the field region. These effects are not taken into account in the above equations. Hence, we have performed simulations to understand the spatial gradient in both electric and magnetic fields. The simulations are also necessary to optimize the field length and field-free drift region for detecting ion energy range as required during the experiment.

\subsection{Simulation of electric and magnetic field}
The electric and magnetic fields have been simulated in COMSOL Multiphysics \cite{comsol} software to understand their spatial distribution. A three-dimensional geometry has been simulated in COMSOL, similar to Fig. \ref{fig:fig1}. The permanent magnets and electrodes have been placed at 60 mm and 18 mm apart, respectively. The electrodes and magnets are of the same length i.e. 150 mm. The remnant magnet flux density assigned to the magnets is 1.3 T. The electrodes have been biased at ±2350 V, respectively, to obtain a potential difference of 4700 V across the field region. The magnetic field has been defined in the direction parallel to the electric field.

\begin{figure}
\includegraphics[scale=0.55]{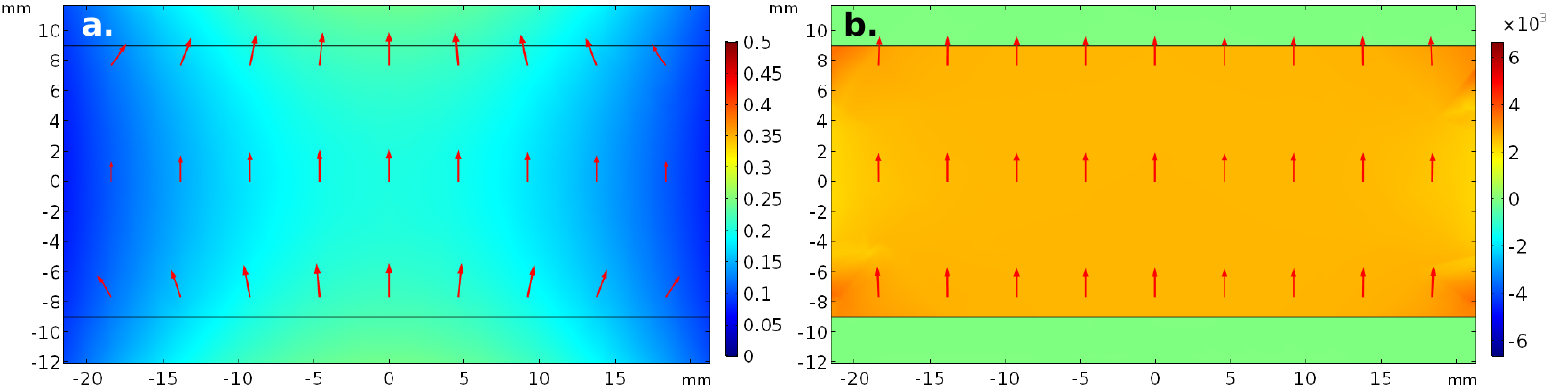}
\caption{\label{fig:fig2} Two-dimensional color map in the XY plane representing \emph{a.} magnetic field and \emph{b.} electric field. The arrows represent the direction of field and its strength.}
\end{figure}

The color maps depicting magnetic and electric field have been shown in Fig. \ref{fig:fig2}. The figure shows the field distribution on the XY plane situated at the centre of field region. The arrows signify the direction of fields and their length depict the field intensity. In this configuration, it has been observed that the electric field as well as magnetic field exhibit good uniformity. The electric and magnetic fields along the spectrometer axis ($z$ axis in Fig. \ref{fig:fig1}) are 2610 V/cm and 0.2 T, respectively.

\begin{figure}
\includegraphics[scale=0.8]{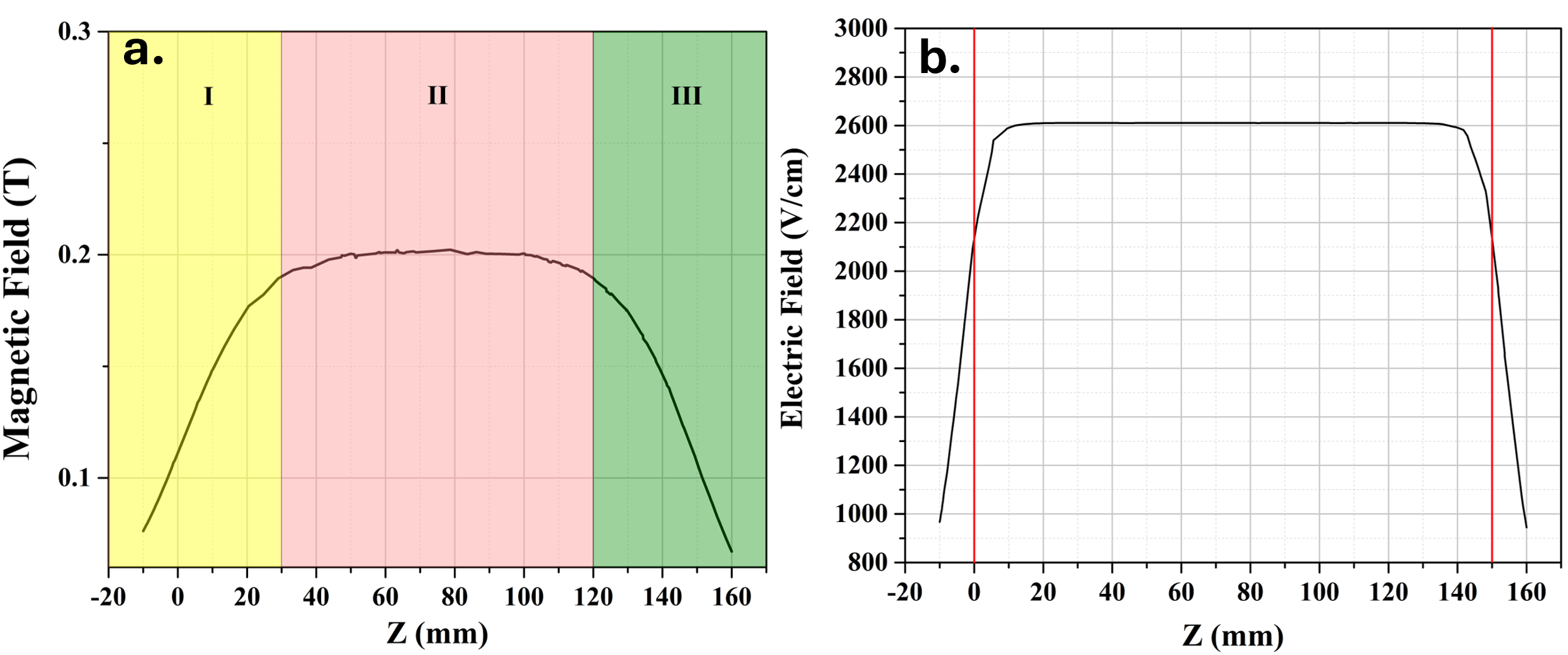}
\caption{\label{fig:fig3} The intensity of $a.$ magnetic field and $b.$ electric field along the ion propagation axis (Z-axis). The colored bands in $a.$ represent different regions in the field profile and the region between red lines denotes the field region length in $b.$}
\end{figure}

We map the field intensity along the spectrometer axis, as shown in Fig \ref{fig:fig3}. The fields remain uniform through most of the ion traversing length in the field region. The gradual rise and fall of fields in the entrance and exit of the field region correspond to the fringe effect due to finite dimensions of electrodes and magnets.

\subsection{Ion trajectory analysis}
\label{sec:ion_traj}
Following the simulation of electric and magnetic field distributions, numerical ion trajectory calculations have been performed to predict their positions on the detector plane. This analysis can help in estimating the minimum ion energy that can be measured in present configuration. As discussed earlier, equations \ref{eqn:xdef0} and \ref{eqn:ydef0} do not account for non-constant magnetic and electric fields. Hence, the following equations have been used for $x$ and $y$ deflections considering a non-uniform electromagnetic field as shown in Fig. \ref{fig:fig3}.

\begin{eqnarray}
\frac{d^2x}{dt^2} = \frac{-qv_zB(z)}{m}
\label{eqn:xdef1}
\end{eqnarray}

\begin{eqnarray}
\frac{d^2y}{dt^2} = \frac{qE(z)}{m}
\label{eqn:ydef1}
\end{eqnarray}
Here, $v_z$ is the velocity of ion along z-axis, $B(z)$ and $E(z)$ are magnetic and electric fields as a function of propagation length. Assuming change in $v_z$ is very small as the ion moves through the field region, the above equations can be written as,

\begin{eqnarray}
\frac{d^2x}{dz^2} = \frac{-qB(z)}{mv_z}
\label{eqn:xdef2}
\end{eqnarray}

\begin{eqnarray}
\frac{d^2y}{dz^2} = \frac{qE(z)}{mv_z^2}
\label{eqn:ydef2}
\end{eqnarray}

\begin{figure}
    \centering
    \includegraphics[width=0.9\linewidth]{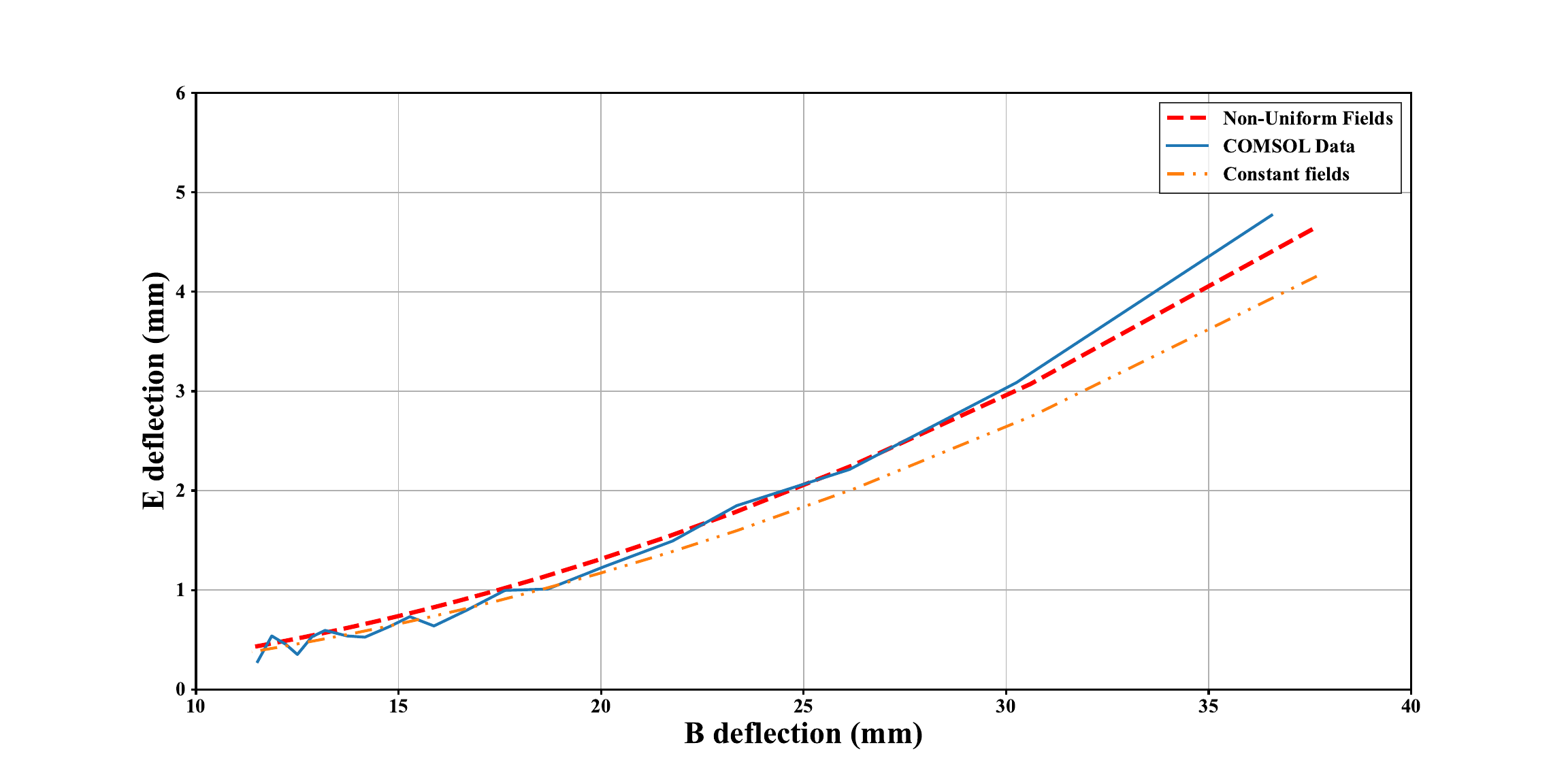}
    \caption{Proton trajectories on the detector plane simulated from the COMSOL data (line), constant field approximation (dot dash line), and non-uniform field (dashed line). The non-uniform field trajectories are in close agreement with COMSOL simulated proton trajectories.}
    \label{fig:theory}
\end{figure}

Assuming magnetic and electric field profiles as shown in Fig. \ref{fig:fig3}, the field profiles are divided into three regions, viz. gradual rise, constant plateau region, and gradual fall. By fitting linear equations to the first and third regions, the following analytical expression is formulated.

\begin{eqnarray}
B(z) = \begin{cases}
B_0[a+bz] \quad \text{for } 0<z<l_1 \\
B_0 \quad \text{for } l_1<z<l_2 \\
B_0[c+dz] \quad \text{for } l_2<z<l_3
\label{eqn:mag1}
\end{cases}
\end{eqnarray}

\begin{eqnarray}
E(z) = \begin{cases}
E_0[a'+b'z] \quad \text{for } 0<z<l'_1 \\
E_0 \quad \text{for } l'_1<z<l'_2 \\
E_0[c'+d'z] \quad \text{for } l'_2<z<l'_3
\label{eqn:elec1}
\end{cases}
\end{eqnarray}
Here, $E_0$ and $B_0$ are the field intensities in the constant plateau region, $l_i$ and $l _i'$ are limits of the three regions depicted by the color bands in Fig. \ref{fig:fig3}a for magnetic and electric field, respectively, $a$, $a'$, $c$, $c'$ are the intercepts for non-uniform field regions, and $b$, $b'$, $d$, $d'$ are the corresponding slopes. 
After fitting equations \ref{eqn:mag1} and \ref{eqn:elec1} with the simulated field profiles, the values of slopes have been obtained and the estimated field profile of $E(z)$ and $B(z)$ are substituted in equation \ref{eqn:xdef2} and \ref{eqn:ydef2} to obtain $x$ and $y$ deflections for a particular ion species. Since the detector has a diameter of 75 mm, the ion trajectory calculations have been performed only till the detector boundary, thereby defining the minimum detectable ion energy. This analysis has been done to consider realistic field profiles which are non-uniform in nature. The comparison between proton trajectories simulated from equations \ref{eqn:xdef0}-\ref{eqn:ydef0} and \ref{eqn:xdef2}-\ref{eqn:ydef2} has been shown in Fig. \ref{fig:theory} with the COMSOL simulated proton trajectory. The proton trajectories simulated by non-uniform fields is in close agreement with the COMSOL simulations. However, the trajectory simulated by considering constant field values underestimates the deflections. The deviation from COMSOL trajectory becomes prominent at lower energies, highlighting the influence of fringe fields and non-uniformities on proton trajectory. 

\begin{figure}
\includegraphics[scale=0.5]{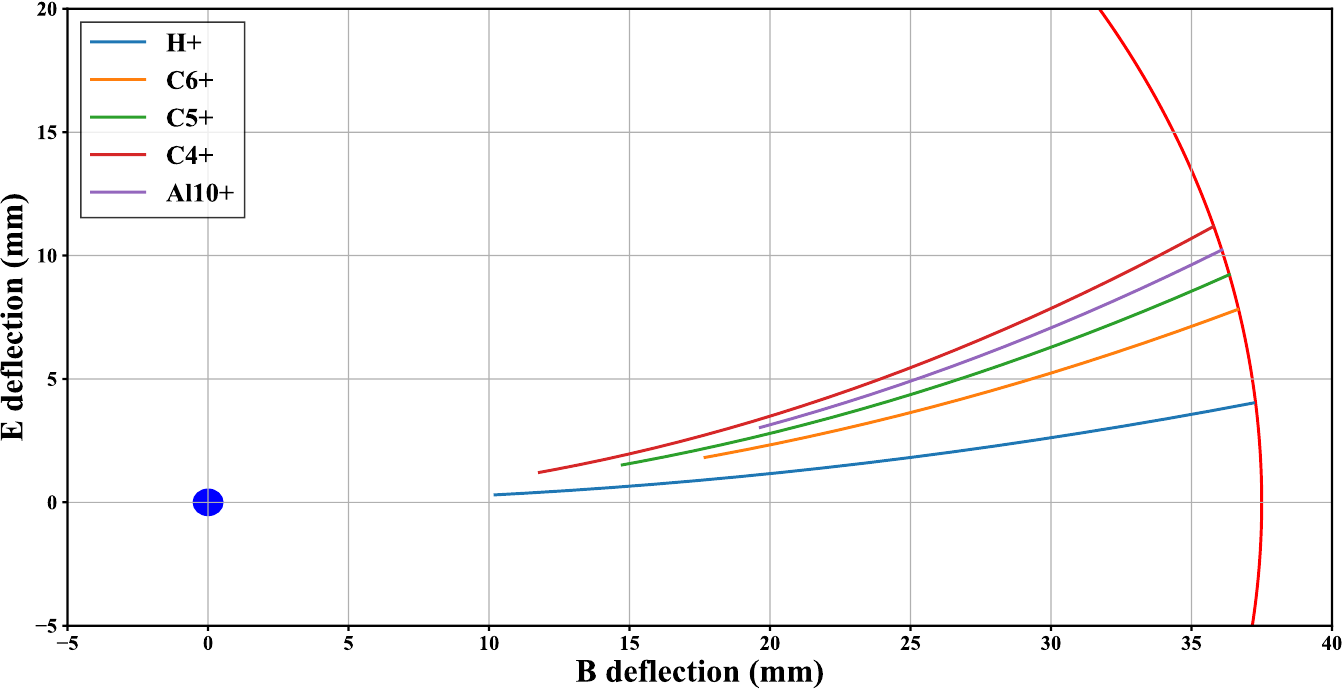}
\caption{\label{fig:fig4} Simulated ion Trajectories for ion species H$^+$, C$^{6+}$, C$^{5+}$, C$^{4+}$, and Al$^{10+}$ with field length, $L_d$ = 80 mm. The limiting curve (red) represents the extent of detector.}
\end{figure}

Ion trajectories analysis has been conducted using deflections obtained from equations \ref{eqn:xdef2} - \ref{eqn:ydef2} at different $L_d$ to optimize the drift length. While a shorter drift length increases the measurable energy range, it also reduces the spatial separation between neighboring ion trajectories, thereby degrading the energy resolution.
A representation of the TPIS ion trajectories have been shown in Fig. \ref{fig:fig4} for $L_d$ = 80 mm. The blue spot denotes the origin of the detector as well as the point where undeflected particles such as X-rays or neutral atoms would be detected as they would be not be affected by the fields of the spectrometer. Traces of different ion species such as H$^+$, C$^{6+}$, C$^{5+}$, C$^{4+}$, and Al$^{10+}$ can be observed. Lighter ions such as protons exhibit comparatively smaller electric deflections (y-deflection) owing to their larger velocities for a given kinetic energy, whereas heavier ions with different charge states occupy distinct regions of the detector plane. The clear spatial separation between the trajectories of  H$^+$, C$^{6+}$, C$^{5+}$, C$^{4+}$, and Al$^{10+}$ demonstrates that the selected field configuration and detector geometry are sufficient to distinguish these ion species without significant overlap.
\begin{table}[ht]
\centering
\caption{Minimum detectable ion energy for different values of drift length ($L_d$).}
\label{tab:tab1}
\begin{tabular}{|c|c|c|c|c|}
\hline
\textbf{Ion species} &
\multicolumn{4}{c|}{\textbf{Minimum detectable ion energy (MeV)}} \\ \cline{2-5}
 & \textbf{60 mm} & \textbf{80 mm} & \textbf{100 mm} & \textbf{120 mm} \\
\hline
H$^{+}$      & 0.57 & 0.75 & 0.95 & 1.18 \\
C$^{6+}$     & 1.79 & 2.33 & 2.94 & 3.62 \\
C$^{5+}$     & 1.27 & 1.64 & 2.07 & 2.54 \\
C$^{4+}$     & 0.84 & 1.08 & 1.36 & 1.66 \\
Al$^{10+}$   & 2.30 & 2.96 & 3.72 & 4.57 \\
\hline
\end{tabular}
\end{table}
Similar analysis has been carried out for different $L_d$ viz. 60, 80, 100, and 120 mm. The minimum detectable ion energy for different ion species is summarized in table \ref{tab:tab1}. Among the ion species considered, $H^+$ exhibits the lowest minimum detectable energy owing to its smaller electric deflection, whereas Al$^{10+}$ requires the highest minimum energy to remain within the detector boundary. For example, the minimum detectable energy of $H^+$ increases from 0.57 MeV at $L_d$ = 60 mm to 1.18 MeV at $L_d$ = 120 mm, while the corresponding value for Al$^{10+}$ increases from 2.30 MeV to 4.57 MeV. These results clearly demonstrate the trade-off between detector drift length and measurable energy range. Although increasing $L_d$ improves the spatial separation between neighboring ion trajectories and in turn, the energy resolution, it simultaneously reduces the acceptance for low-energy ions. Based on this analysis, a drift length of 80 mm was selected as an optimum compromise between ion separation and the measurable energy range. 

\begin{figure}
\includegraphics[scale=0.5]{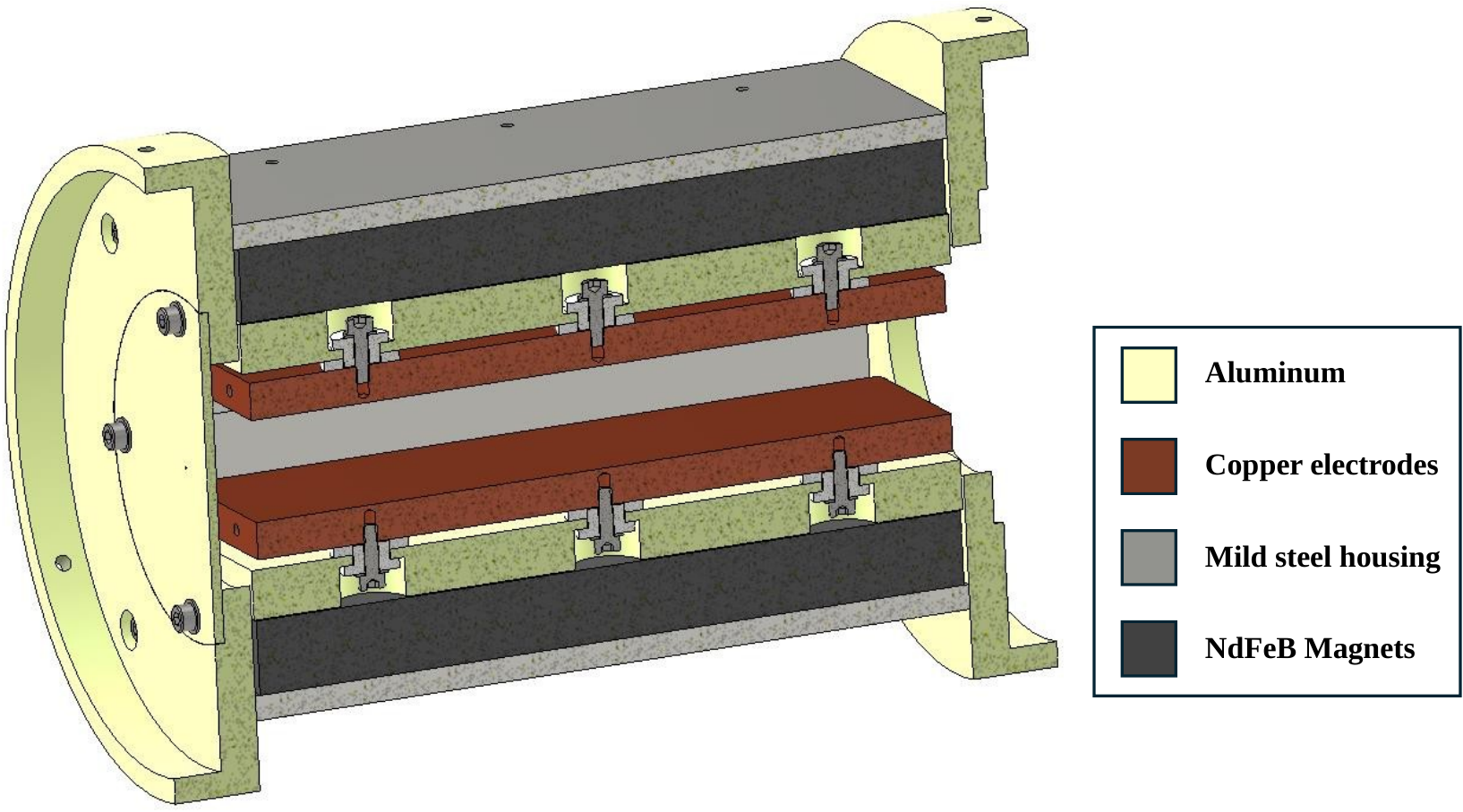}
\caption{\label{fig:fig5} CAD drawing of the fabricated TPIS showing internal details of electrodes and magnets. The colors in the image depict different materials. An aperture at the entrance of the spectrometer can also be seen in this figure.}
\end{figure}

\section{Design of Thomson parabola ion spectrometer}
The TPIS has been fabricated in-house at the PRL workshop based on the optimized design obtained through numerical analysis and finite-element simulations of the electric and magnetic field distributions. Fig. \ref{fig:fig5} shows the CAD drawing of fabricated TPIS and its major components. The spectrometer consists of an entrance aperture, an electrostatic deflection assembly, a permanent magnet assembly, and a detector positioned after the drift region. The entrance aperture samples only the collimated portion of the incoming ion beam and minimizes background arising from ion cloud propagating outside the central propagation axis. The aperture assembly consists of a 2 mm thick lead (Pb) disk with 0.25 mm aperture, sandwiched between 2 mm thick aluminum (Al) disks which have 1 mm diameter aperture. Pb aperture is utilized to suppress the intense X-ray background generated by laser produced plasma.

\begin{figure}
\includegraphics[scale=0.5]{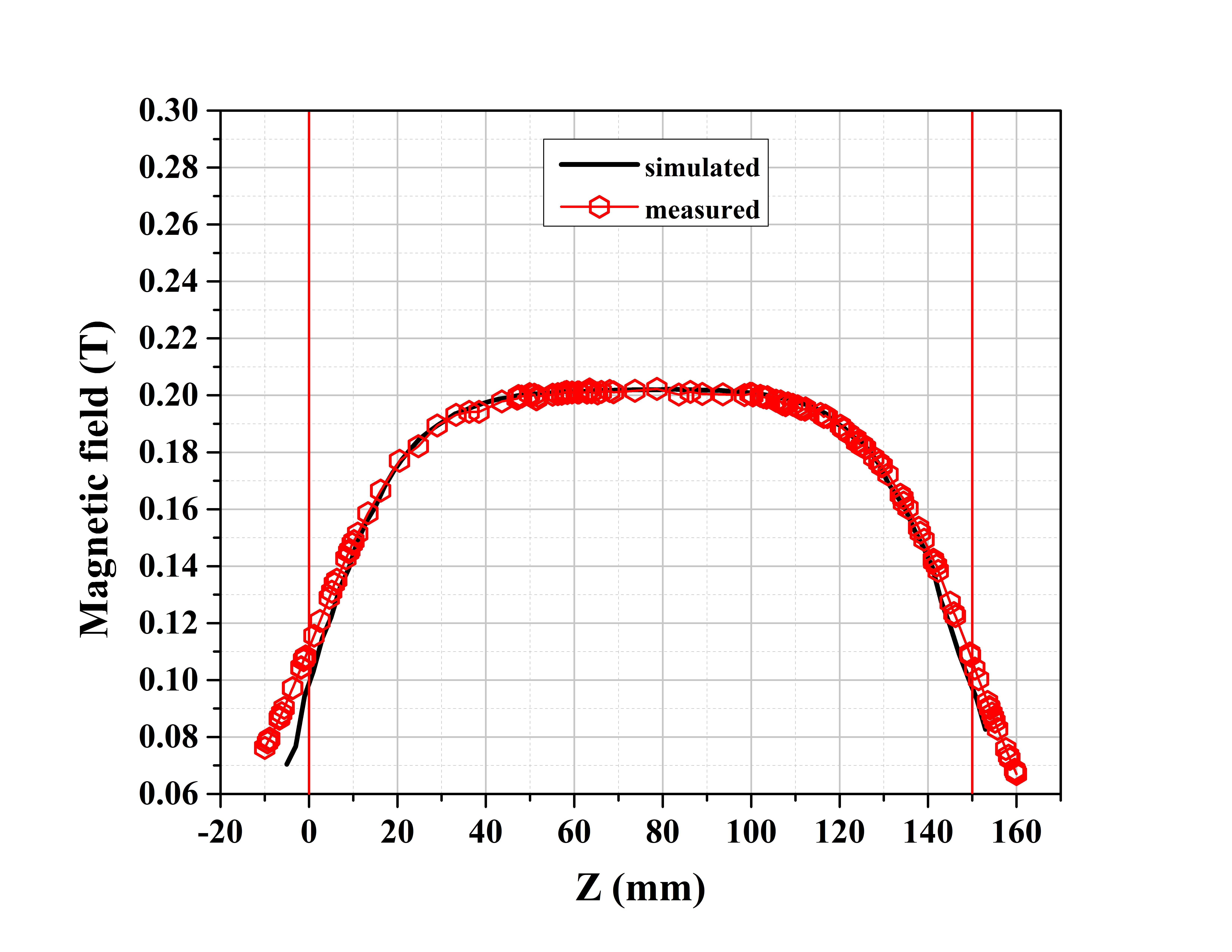}
\caption{\label{fig:fig6} A comparison between simulated (black line) and experimentally measured magnetic field (red symbol) along z-axis.}
\end{figure}

The electrostatic and magnetic deflection sections are mounted coaxially to ensure that ions experience mutually perpendicular electric and magnetic fields during propagation and to make the setup more compact. The electrostatic plates are mounted on an Al plate and isolated using teflon spacers to maintain a constant electrode separation while withstanding the applied high voltage. The assembly of electrodes and permanent magnets has been grounded to prevent stray voltage that could perturb ion trajectories. The NdFeB magnets have a pole strength of 1.3 T and provides a magnetic field of 0.2 T along the spectrometer axis. The magnetic field has been measured experimentally using gauss-meter, along the spectrometer axis and is shown in Fig. \ref{fig:fig6}. The measured magnetic field profile is in excellent agreement with the simulated values, depicting the accurate realization of the simulated magnetic field. The normalized root mean square error (NRMSE) between the simulated and measured magnetic field is 1.75\%, while the $R^2$ value is 0.9870 in the field length region (between the red lines). The close agreement between the measured and simulated field profiles validates the numerical model employed during the design stage. Minor deviations observed near the edges of the magnetic field region can be attributed to machining tolerances, and positioning error during gauss meter measurements.

\begin{figure}
\includegraphics[scale=0.5]{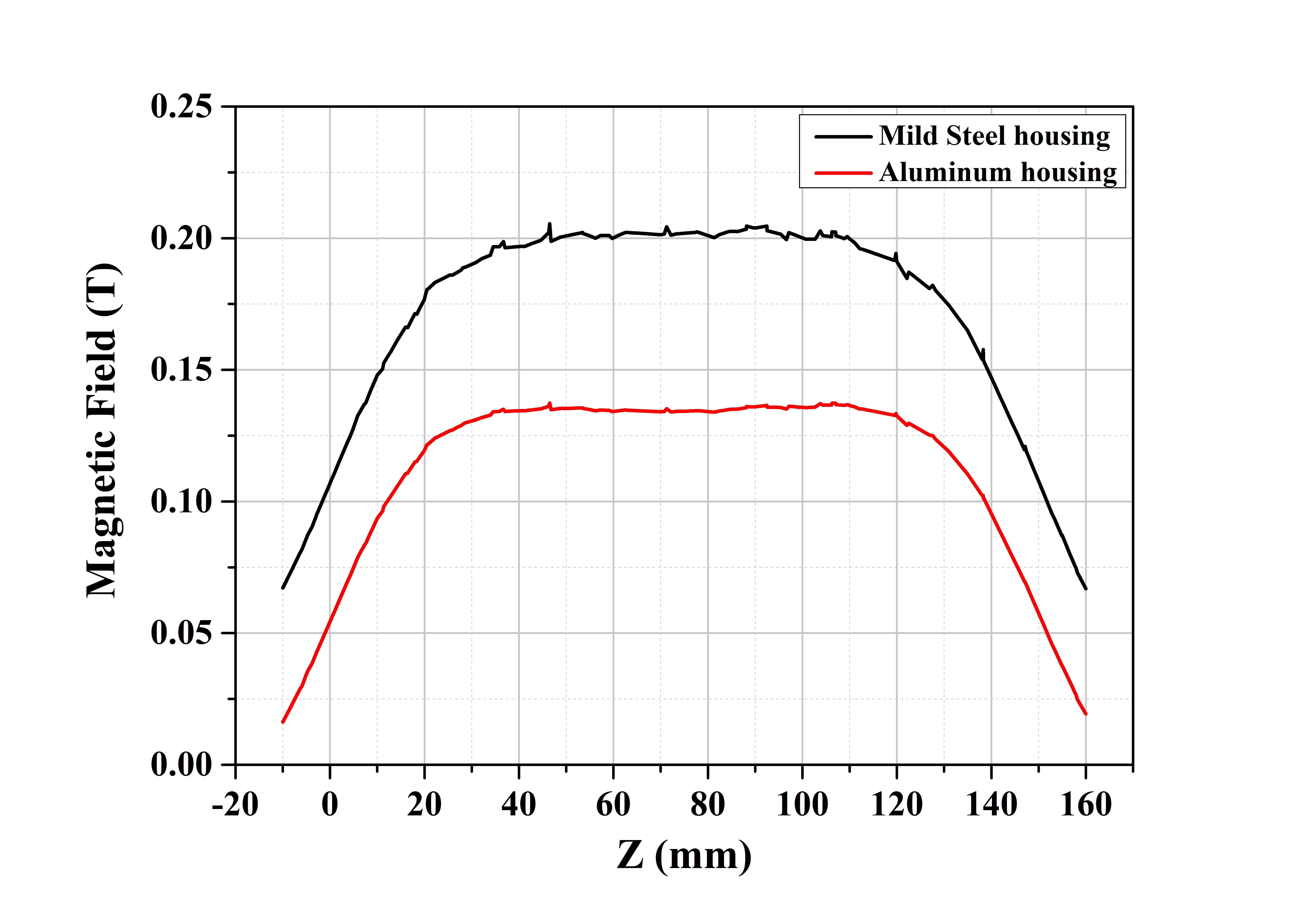}
\caption{\label{fig:fig7} Magnetic field for TPIS magnets and electrodes housed in
MS enclosure (black) and aluminum enclosure (red).}
\end{figure}

The magnets and electrodes are housed within a mild-steel (MS) yoke, which serves as a magnetic return path and suppresses magnetic flux leakage into the surrounding region. To quantify the influence of housing material, the magnetic field along spectrometer axis has been mapped in COMSOL with aluminum as housing material and later with MS, as shown in Fig. \ref{fig:fig7}. The magnetic field is 0.13 T with aluminum housing. These results indicate that magnets enclosed in MS housing exhibits significant enhancement in magnetic field.

The detector assembly consists of a chevron micro-channel plate (MCP) with active diameter of 75 mm coupled to a phosphor screen, enabling single-shot detection of laser- accelerated ions. The detector position corresponds to the optimum drift length obtained from trajectory optimization described in section \ref{sec:ion_traj} to provide an appropriate compromise between measurable energy range and energy resolution.

\begin{figure}
\includegraphics[scale=0.1]{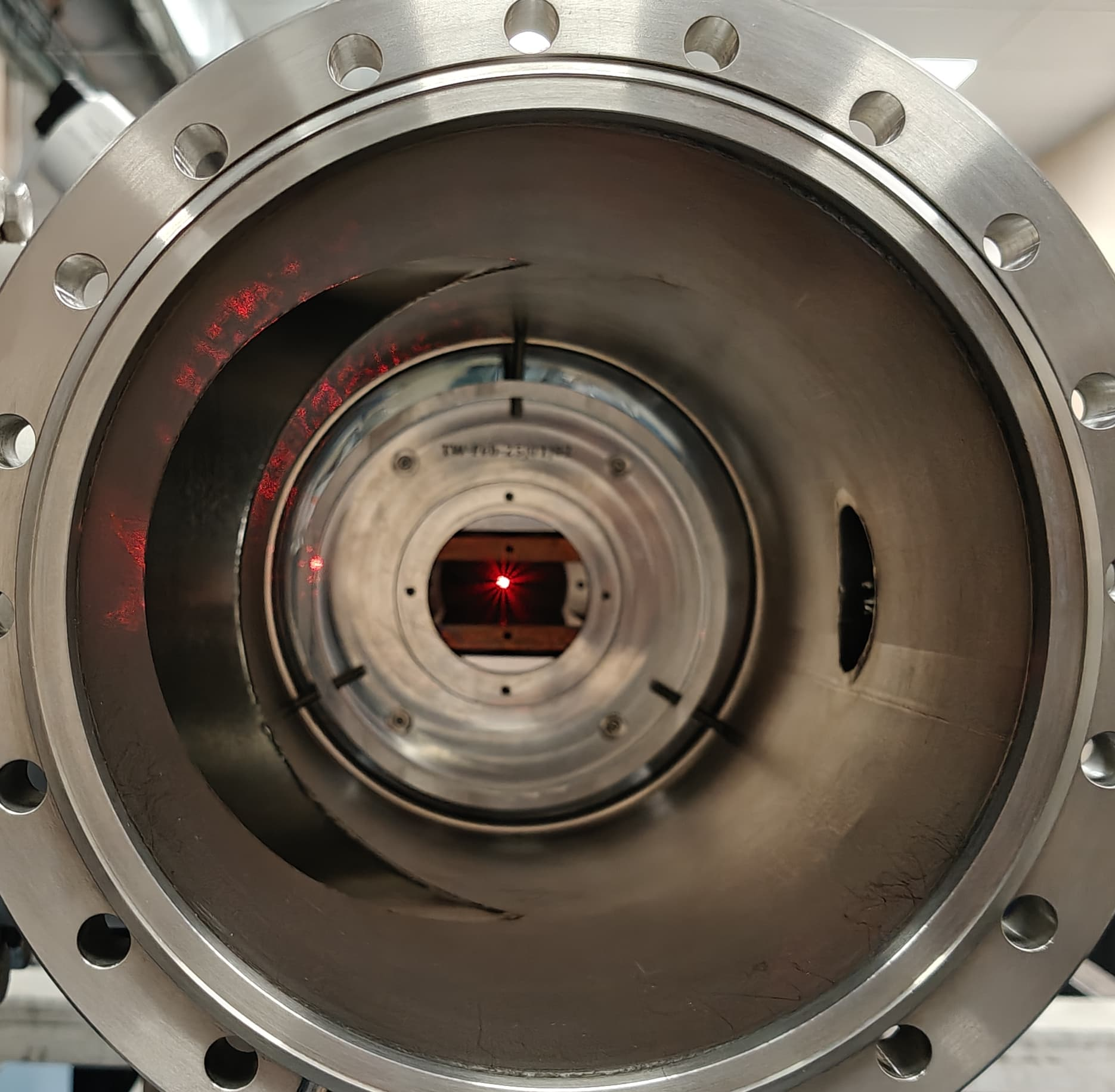}
\caption{\label{fig:fig8} The fabricated TPIS mounted inside a vacuum chamber. A He-Ne laser directed along the aperture is used for alignment of TPIS inside the chamber.}
\end{figure}

Fig. \ref{fig:fig8} shows the fabricated TPIS integrated within a vacuum chamber. The spectrometer has been mounted on circular support flanges on both the sides. These flanges are provided with axial tapping at 120$^\circ$ interval ensuring precise alignment of the aperture, deflection assemblies, and detector along the spectrometer axis to minimize systematic errors in ion trajectory measurements. A notable feature of the present design is the provision for translating the spectrometer within vacuum chamber. This allows to tune the drift length ($L_d$) without modifying the geometry of spectrometer. Consequently, the measurable ion energy range can be tailored to different experimental requirements. This flexibility enables the same spectrometer to be employed for experiments involving different ion species and energy ranges without requiring any modification to the electrostatic or magnetic deflection assemblies. 

The final geometrical dimensions and operating parameters of the developed TPIS are summarized in Table \ref{tab:tab2}.
\begin{table}
\caption{\label{tab:tab2}Final specification of the designed TPIS.}
\begin{ruledtabular}
\begin{tabular}{|c|c|}
\textbf{Parameter} & \textbf{Value}\\
\hline
Entrance aperture diameter & 0.25 mm\\ \hline
Electrode length & 150 mm\\\hline
Gap between electrodes & 18 mm\\\hline
Maximum voltage across electrodes & 4700 V\\\hline
Maximum Electric field & 2610 V/cm\\\hline
Magnet material & NdFeB \\\hline
Magnet length & 150 mm \\\hline
Magnet pole strength & 1.3 T \\\hline
Magnetic field along z-axis & 0.2 T\\\hline
Drift length & 80 mm (tunable)\\\hline
Detector & Chevron MCP (75 mm) + Phosphor Screen
\end{tabular}
\end{ruledtabular}
\end{table}

\section{Experimental validation}
The realized TPIS has been tested during experimental campaign at the petawatt laser facility of RRCAT, Indore, India. The aim of this experiment has been to demonstrate the capability of the spectrometer during laser-driven ion acceleration experiments. The experimental configuration and representative ion traces obtained using the TPIS are presented in this section.

\subsection{Setup}

\begin{figure}
\includegraphics[scale=0.7]{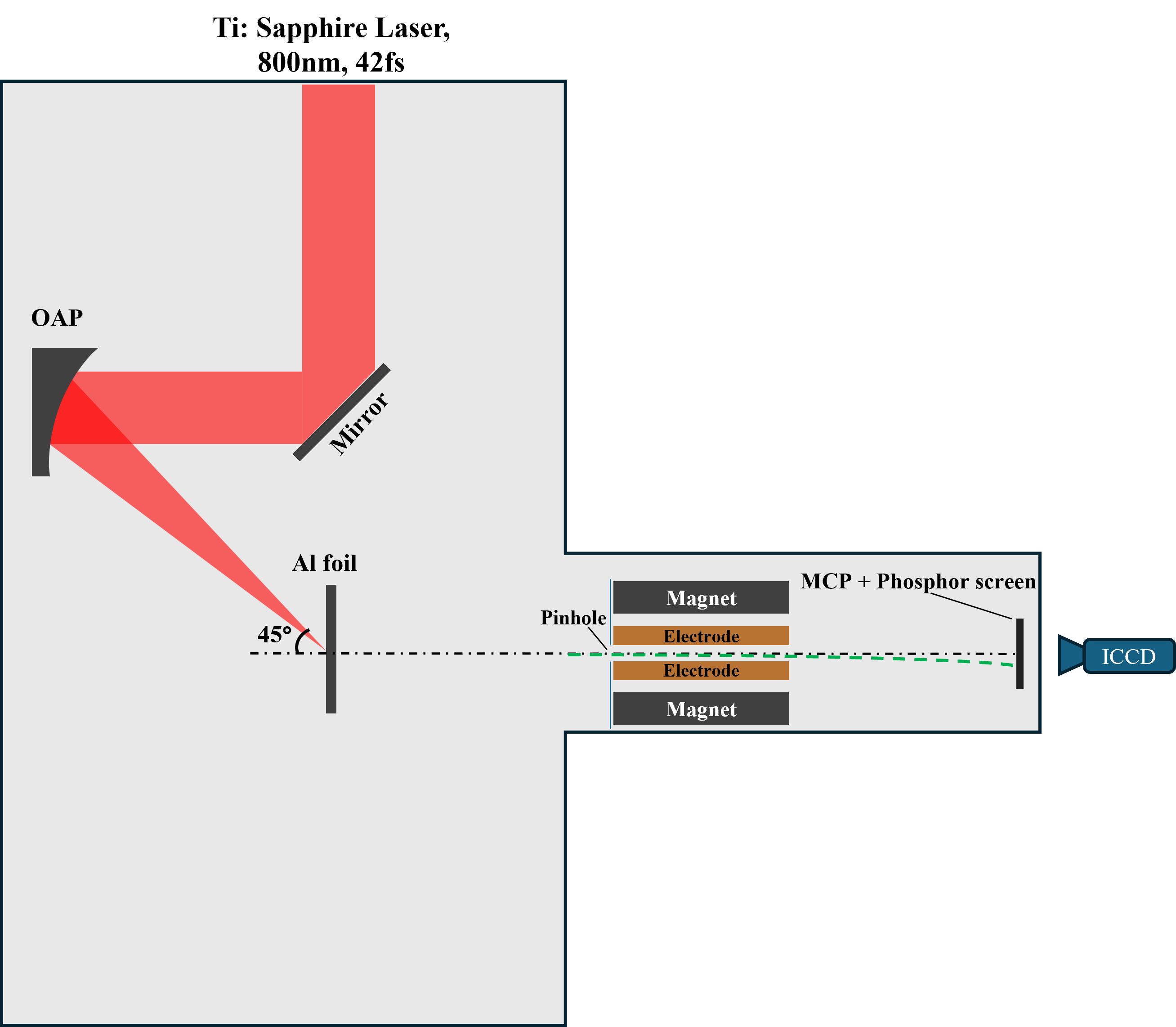}
\caption{\label{fig:fig9} Schematic of the experimental setup to test the TPIS during laser-driven ion acceleration campaign at the petawatt laser facility, RRCAT, Indore.}
\end{figure}

The schematic of the experimental arrangement to test the TPIS is shown in Fig. \ref{fig:fig9}. The experiments have been carried out using Ti: Sapphire laser system (Thales, $\lambda$ = 800 nm) delivering pulses of 42 fs duration. The pulse energy on the target is 0.8 J and the laser has a pre-pulse contrast of 10$^{-10}$. The pulse is focused using an off-axis parabolic (OAP) mirror of 600 mm effective focal length to a spot of $\sim$ 5 $\mu$m to achieve a peak intensity of $6.8 \times 10^{19}$ W/cm$^2$. Two targets have been utilized for the experiments viz., 0.8 $\mu$m and 10 $\mu$m thick Al foil mounted on a specialized target mount. The incidence angle of laser with respect to the target normal is 45$^\circ$. The aperture of TPIS is located at a distance of $\sim$1.2 m from the target, allowing only the collimated ion beam into the deflection region. The detector assembly consists of MCP + phosphor screen and is placed 80 mm away from the field region. The emission from phosphor screen is captured by an intensified charge couple device (ICCD) camera placed outside the vacuum chamber. The ICCD camera is externally triggerable and synchronized with the laser pulse. The interaction chamber is maintained at a base pressure of $10^{-6}$ mbar while the chamber with TPIS is differentially pumped and maintained at $10^{-7}$ mbar. 

\subsection{Data extraction and analysis}
\begin{figure}
\includegraphics[scale=0.5]{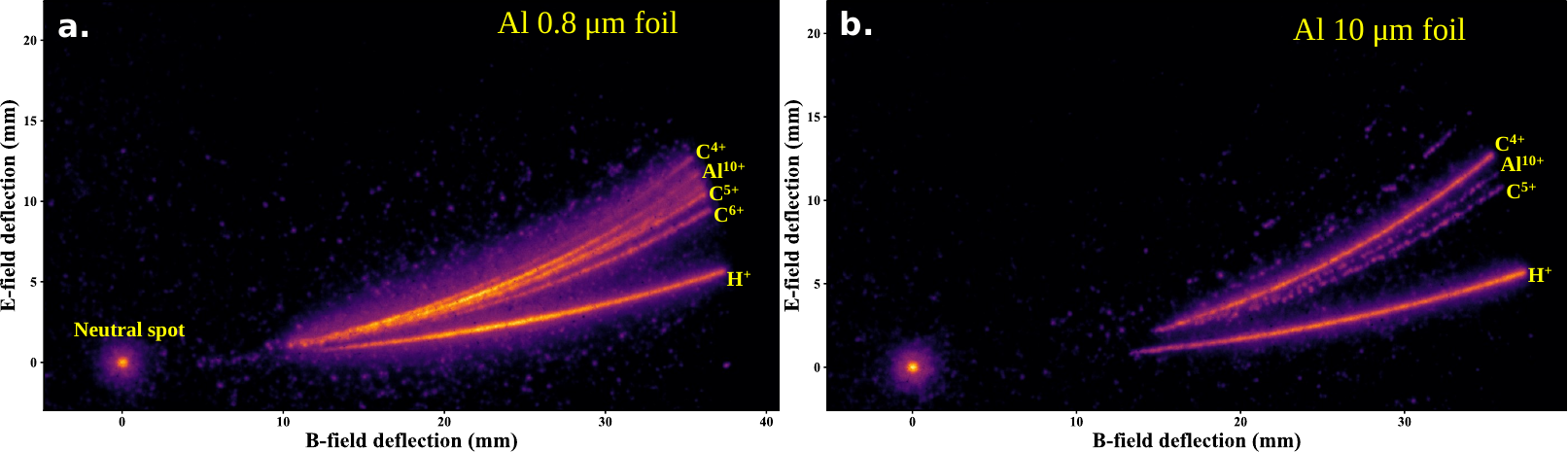}
\caption{\label{fig:fig10} Raw ion traces from ion acceleration experiment with target $a.$ 0.8 $\mu$m and $b.$ 10 $\mu$m Al foil. The ion species and neutral spot are labeled in the image (yellow).}
\end{figure}

The detector images of the developed TPIS as a result of laser interaction with 0.8 $\mu$m and 10 $\mu$m thick Al foils are presented in Fig. \ref{fig:fig10}. The bright spot located near the lower left corner is the neutral spot. Traces from ion species H$^+$, C$^{4+}$, C$^{5+}$, C$^{6+}$, Al$^{10+}$ have been observed and were resolved by the TPIS during these experiments, demonstrating its capabilities.

\begin{figure}
\includegraphics[scale=0.5]{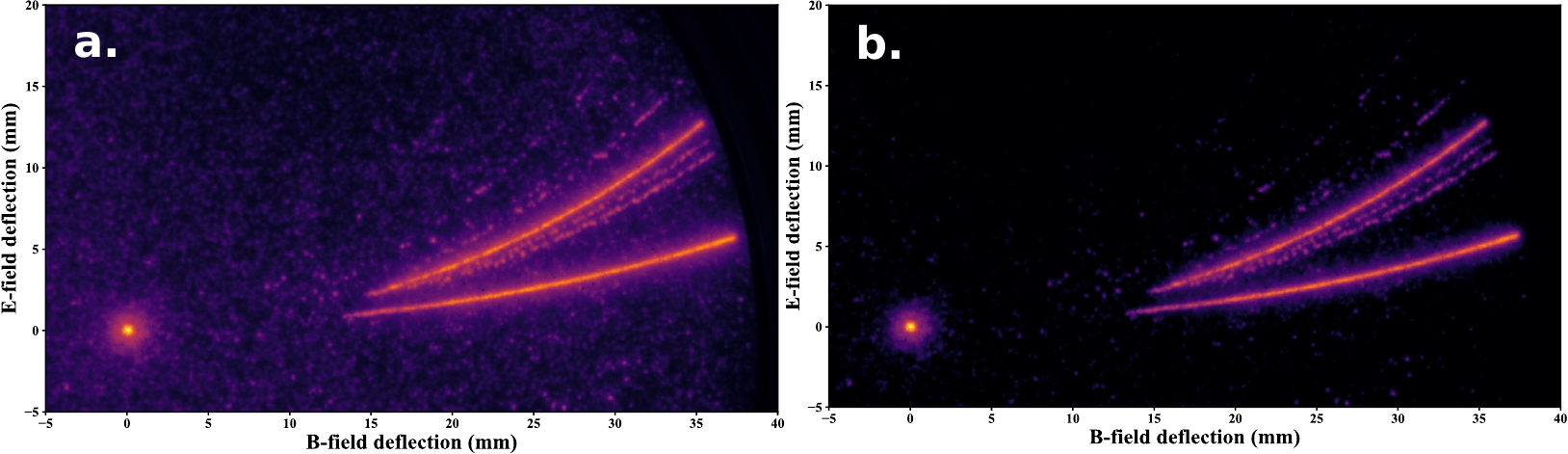}
\caption{\label{fig:fig11} TPIS images used for processing. $a.$ Unprocessed image $b.$ Background subtracted image. The contrast between background and the ion traces has improved after background subtraction.}
\end{figure}

\begin{figure}
\includegraphics[scale=0.4]{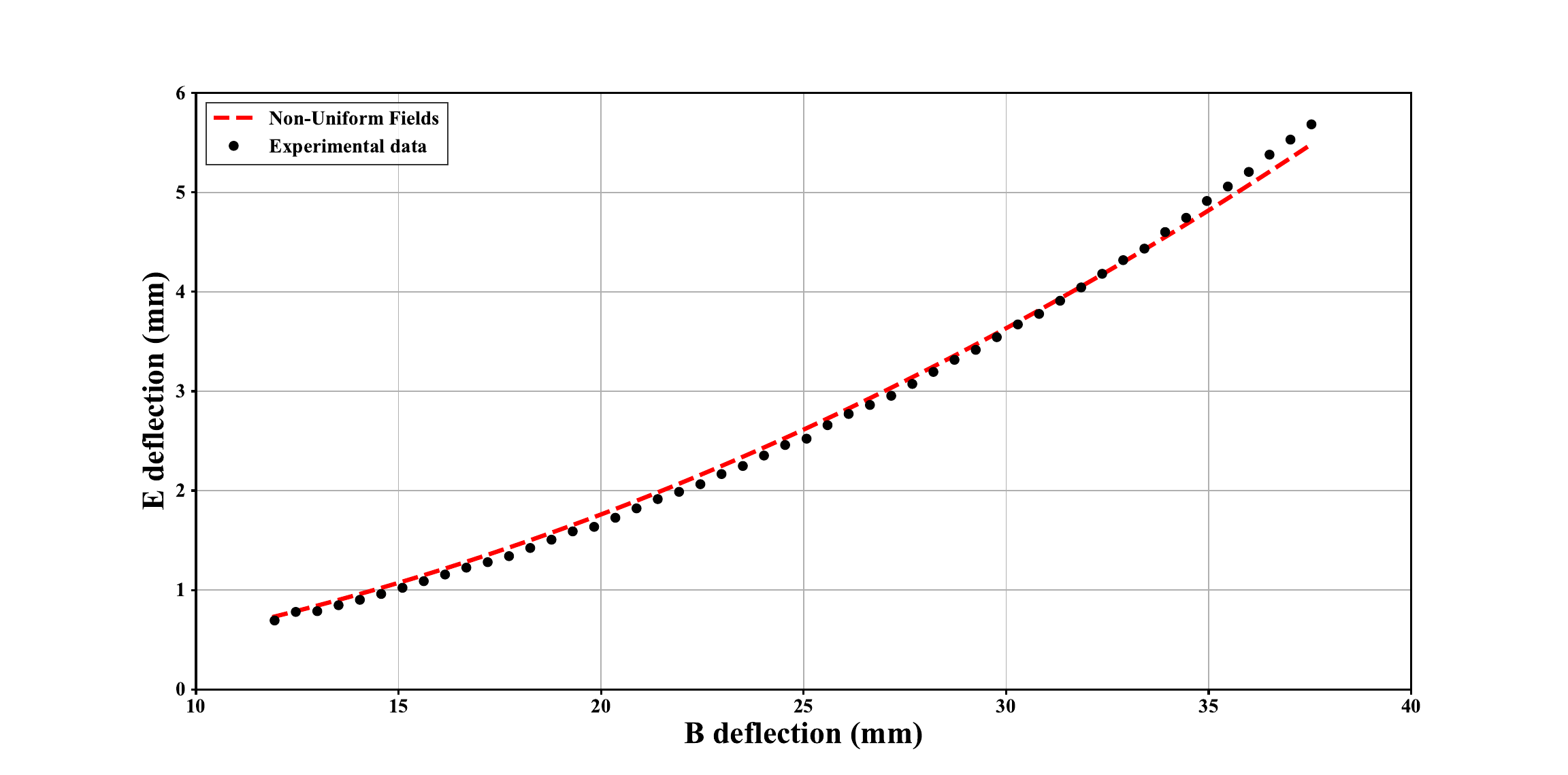}
\caption{\label{fig:fig12} Fitting of experimental proton trajectory (black dot) with the analytical equation accommodating simulated non-uniform field profiles (dashed line).}
\end{figure}

A Python program has been developed to extract the ion energy spectra from the parabolic traces captured by the detector. The deflection of an energetic ion is determined by the electric and magnetic fields acting on it as given by equations \ref{eqn:xdef2}-\ref{eqn:ydef2}. The neutral spot is first identified to establish the coordinate system, followed by background subtraction (Fig. \ref{fig:fig11}). The theoretical parabolic traces generated from the TPIS equations \ref{eqn:xdef2}-\ref{eqn:ydef2} are overlaid on the ion traces obtained after proper coordinate transformation. As discussed before, equations \ref{eqn:xdef2}-\ref{eqn:ydef2} take into consideration a non-uniform electric and magnetic field. Fig. \ref{fig:fig12} is a plot which shows the comparison of proton trajectory simulated by spatially varying fields with the experimentally obtained data. The trajectory simulated incorporating non-uniform field distribution shows excellent agreement with experimental data over the entire deflection range. The experiment trajectory also closely matches the proton trajectory simulated by equations \ref{eqn:xdef0}-\ref{eqn:ydef0} for this case. In addition to using analytical expression there is a minor mismatch between the simulated and experimental parabola which arises due to minor misalignment during the experiment. The misalignment may arise due to minor mismatch between spectrometer axis and ion propagation axis during coupling of TPIS chamber with the interaction chamber. This mismatch has been corrected by rotating the image with a small angle ($\sim$1$^\circ$) with origin as the center of rotation.

\begin{figure}
\includegraphics[scale=0.5]{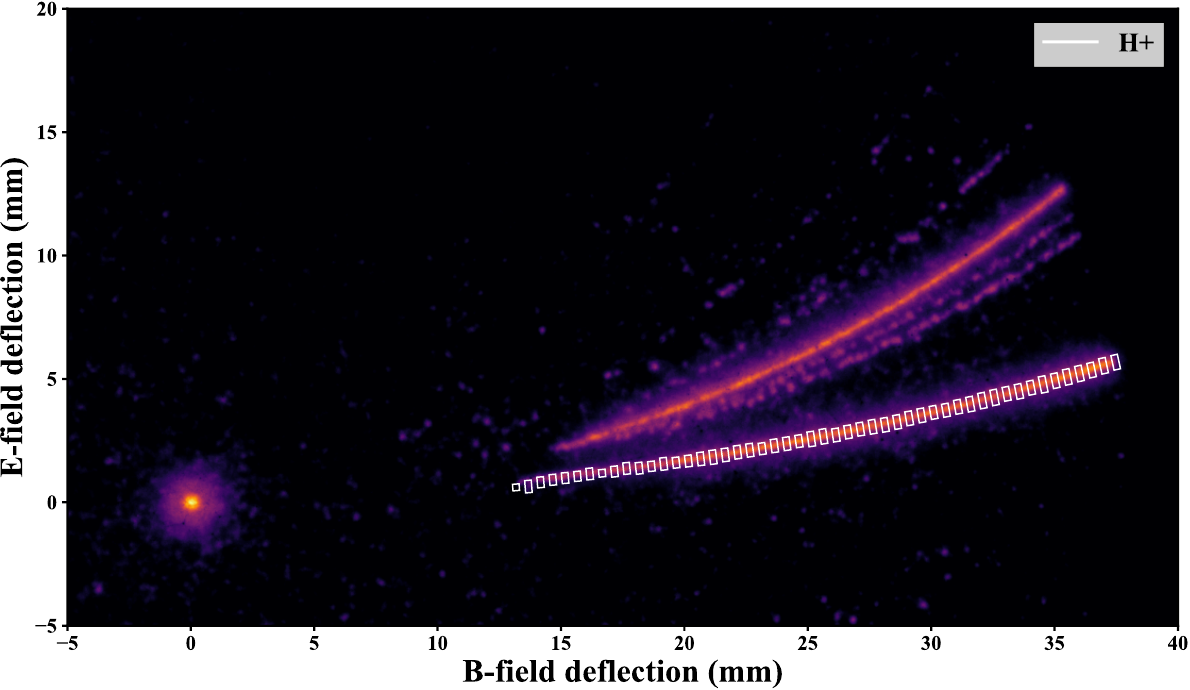}
\caption{\label{fig:fig13} Theoretically computed proton trajectory overlaid on the TPIS image to extract the proton energy spectrum.}
\end{figure}

The ion signal is sampled along each ion trajectory using a self-corrected box technique, in which the sampling boxes follow the experimental ion trace as shown in Fig. \ref{fig:fig13}. The average pixel intensity within each box is assigned to the corresponding energy interval, and the energy spectrum is obtained by plotting $dN/dE$ ($\sim \frac{N}{\Delta E}$). The analysis program simultaneously reconstructs the energy spectra of all ion species recorded in the detector image. 

The energy resolution of TPIS has been evaluated from the trajectory of proton in Fig. \ref{fig:fig10}. The energy resolution $\frac{\Delta E}{E}$ for non-relativistic case is given as \cite{Schillaci_2014},

\begin{eqnarray}
\frac{\Delta E}{E}=\frac{2\Delta y}{qBL_f(L_d + 0.5\times L_f)}\times \sqrt{2mE}
\end{eqnarray}
where, $\Delta y$ is the ion trace width, $B$ is the magnetic field, $q$ is the charge state, $m$ is the mass, and $E$ is the kinetic energy of the ion. The developed TPIS has an energy resolution of 2.5\% for 1 MeV protons. The resolution is sufficient to resolve the broad energy range of protons detected in the experimental campaign. The clear separation of the proton trace on the detector (Fig. \ref{fig:fig10}-\ref{fig:fig12}),
together with the agreement between the experimentally observed trajectories and the numerical calibration, enabled reliable extraction of the proton energy spectrum without ambiguity arising from neighboring ion species.

\begin{figure}
\includegraphics[scale=0.4]{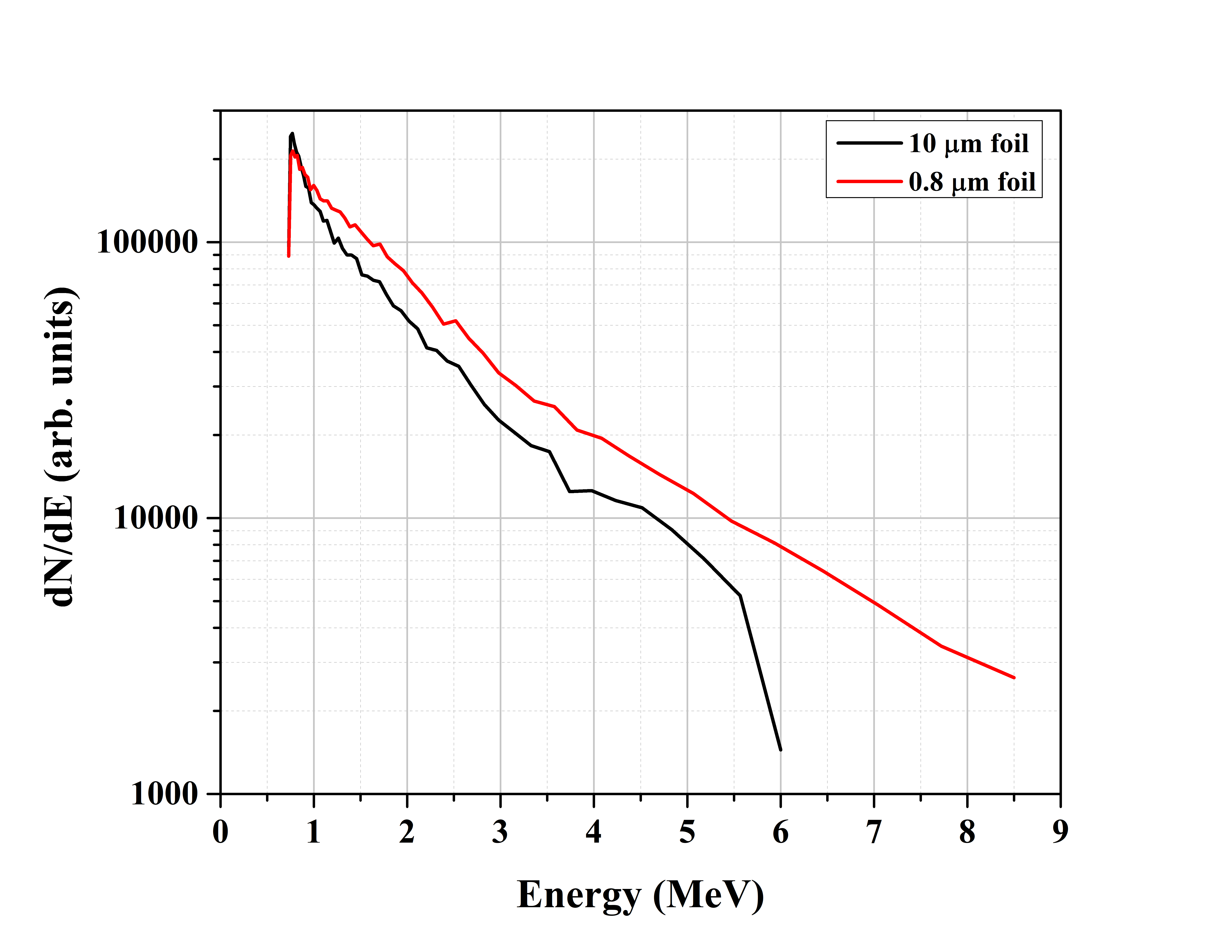}
\caption{\label{fig:fig14} Proton energy spectrum for 0.8 $\mu$m and 10 $\mu$m Al foils. Experiments with 0.8 $\mu$m foil results in higher energy protons and higher proton yield.}
\end{figure}

The proton energy spectrum computed using the aforementioned code is shown in Fig. \ref{fig:fig14} for both the foil targets. It is evident that 0.8 $\mu$m foil has delivered high cutoff energy protons of 8.5 MeV while cutoff energy for 10 $\mu$m foil is 6 MeV. This enhancement can be attributed to reduced hot electron transport distance. The hot electrons have to travel less distance in the bulk foil which reduces dissipation of energy in the surrounding material. The increased hot electron energy density and reduced electron transport distance supports formation of a more localized and prominent sheath field at the rear surface. Moreover, for thin targets, hot electrons can re-circulate between the front and rear sheath fields during the laser pulse, increasing hot electron density, strengthening the sheath field, and consequently enhancing proton acceleration \cite{PhysRevLett.88.215006}. As a result, the ionization of contamination layer and subsequent acceleration of ions in the sheath field results in significantly high energy ions and enhancement in ion yield. The proton cut-off energy also depends on the laser contrast, and efficiency of laser-target coupling.

Similar energy spectra can be obtained from the TPIS image for heavy ion species with accurate numerical fitting using the aforementioned code. The successful reconstruction of proton and heavy-ion energy spectra demonstrates the capability of the developed TPIS for fast diagnostics of multi-species ions  from laser-plasma interactions. Furthermore, the experimentally observed ion trajectories show close agreement with the analytical trajectories obtained using the non-uniform electric and magnetic field profiles. Incorporation of the spatial field variation provides a more accurate energy spectra of the ions than the constant-field approximation. The close match of the experimentally observed parabola trace with the derived analytical expression enable tuning of the measurable energy range of the spectrometer by changing the drift length without the need for recalibration. 

\section{Conclusion}
In this study, we present the design, simulation and development of a Thomson parabola ion spectrometer for diagnostics of laser-accelerated ions. The electric and magnetic field of the spectrometer have been simulated and numerically analyzed using COMSOL Multiphysics to understand the field distribution. Analytical equations incorporating the experiment-validated non-uniform electric and magnetic field profiles have been developed. These equations significantly improved agreement with experiment ion trajectories compared to conventional constant-field approximation. These calculations have also been performed to estimate the minimum measurable ion energy in the active area of the detector for various ion species. Calculations have been performed to optimize the drift length $L_d$ of the spectrometer, and it has been estimated that 80 mm provides a balanced trade-off between detectable ion range and spectrometer resolution and ion species separation. The TPIS has been fabricated in-house with the finalized dimensions and parameters and is characterized experimentally. The measured magnetic field has been observed to be in excellent agreement with the simulated magnetic field. The MS housing encompassing the magnets and electrodes have proven to significantly enhance the magnetic field while improving the field confinement within the deflection region. An additional design feature are the flange which enables $L_d$ adjustment. This feature allows changing the measurable energy range to be tailored as per the experimental need. To validate capabilities of the spectrometer, it has been utilized during experimental campaign at a petawatt laser facility. The spectrometer demonstrated its capability to detect and resolve several ion species including H$^+$, C$^{4+}$, C$^{5+}$, C$^{6+}$, Al$^{10+}$. A robust code has been developed to fit the theoretical ion trace to the corresponding trace in the detector image and construct ion energy spectrum. The close agreement between the experimentally observed ion traces and the predictions of the proposed analytical model confirms the accuracy of the spectrometer design and calibration. Overall, the fabricated TPIS provides a robust and compact diagnostic for laser-plasma experiments.
\section{Acknowledgements}
The authors (K.M, J.P., P.K., H.A.) wish to acknowledge the funding support from Department of Space (DoS), Government of India. The authors also thank the PRL workshop for fabrication of the spectrometer. The authors acknowledge Riyaz Khan and Sunil Kumar Meena for laser operations, as well as S. Sebastin, K. Parmar, and L. Kisku for providing ground support during the experimental campaign. 
\section{Bibliography}
\bibliography{ref}

\end{document}